\documentclass[usegraphicx,usenatbib,useapjfonts,apj]{emulateapj}

\usepackage{apjfonts}
\usepackage{amsbsy}
\usepackage{graphicx}
\usepackage{dcolumn}

\renewcommand\({\left(}
\renewcommand\){\right)}
\renewcommand\[{\left[}
\renewcommand\]{\right]}

\newcommand{\ra}{\rightarrow}

\def\lsim{\raise 0.4ex\hbox{$<$}\kern -0.8em\lower 0.62
ex\hbox{$\sim$}}

\def\gsim{\raise 0.4ex\hbox{$>$}\kern -0.7em\lower 0.62
ex\hbox{$\sim$}}

\def\lbar{{\hbox{$\lambda$}\kern -0.7em\raise 0.6ex
\hbox{$-$}}}

\newcommand\eq[1]{eq.~(\ref{#1})}
\newcommand\eqs[2]{eqs.~(\ref{#1}) and (\ref{#2})}
\newcommand\Eq[1]{Equation~(\ref{#1})}

\newcommand\pa{\partial}
\newcommand\p{\partial}

\newcommand\ee{\end{equation}}
\newcommand\be{\begin{equation}}
\def\bea{\begin{array}}
\def\eea{\end{array}}\def\ea{\end{array}}
\newcommand\ees{\end{eqnarray}}
\newcommand\bees{\begin{eqnarray}}
\newcommand\sub[1]{_{\rm #1}}

\def\p1{{\bf p}_1}
\def\p2{{\bf p}_2}
\def\k1{{\bf k}_1}
\def\k2{{\bf k}_2}

\newcommand{\dddM}{\kern 0.2em \raise 1.9ex\hbox{$...$}\kern -1.0em \hbox{$M$}}
\newcommand{\dddQ}{\kern 0.2em \raise 1.9ex\hbox{$...$}\kern -1.0em \hbox{$Q$}}
\newcommand{\dddI}{\kern 0.2em \raise 1.9ex\hbox{$...$}\kern -1.0em\hbox{$I$}}
\newcommand{\dddJ}{\kern 0.2em \raise 1.9ex\hbox{$...$}\kern-1.0em
\hbox{$J$}}
\newcommand{\dddcalJ}{\kern 0.2em \raise 1.9ex\hbox{$...$}\kern-1.0em
\hbox{${\cal J}$}}

\newcommand{\dddO}{\kern 0.2em \raise 1.9ex\hbox{$...$}\kern -1.0em
\hbox{${\cal O}$}}
\def\dddz{\raise 1.5ex\hbox{$...$}\kern -0.8em \hbox{$z$}}
\def\dddd{\raise 1.8ex\hbox{$...$}\kern -0.8em \hbox{$d$}}
\def\dddbd{\raise 1.8ex\hbox{$...$}\kern -0.8em \hbox{${\bf d}$}}
\def\ddbd{\raise 1.8ex\hbox{$..$}\kern -0.8em \hbox{${\bf d}$}}
\def\dddx{\raise 1.6ex\hbox{$...$}\kern -0.8em \hbox{$x$}}

\newcommand{\msun}{M_{\odot}}

\def\D{\Delta}
\def\p{\partial}
\def\a{\alpha}
\def\b{\beta}

\def\nn{\nonumber}

\def\s{\sigma}
\def\g{\gamma}
\def\G{\Gamma}
\def\d{\delta}

\def\eps{\epsilon}

\def\dslash{\hspace{-1mm}\not{\hbox{\kern-2pt $\partial$}}}
\def\Dslash{\not{\hbox{\kern-4pt $D$}}}
\def\pslash{\not{\hbox{\kern-2.1pt $p$}}}
\def\kslash{\not{\hbox{\kern-2.3pt $k$}}}
\def\qslash{\not{\hbox{\kern-2.3pt $q$}}}

\begin{document}

\title{The Halo Mass Function from Excursion Set Theory.\\
II. The Diffusing  Barrier}

\author{
Michele Maggiore\altaffilmark{1} and
Antonio Riotto\altaffilmark{2,3}
}
\altaffiltext{1}{D\'epartement de Physique Th\'eorique, 
Universit\'e de Gen\`eve, 24 quai Ansermet, CH-1211 Gen\`eve, Switzerland}
\altaffiltext{2}{CERN, PH-TH Division, CH-1211, Gen\`eve 23,  Switzerland}
\altaffiltext{3}{INFN, Sezione di Padova, Via Marzolo 8,
I-35131 Padua, Italy}

\begin{abstract}

In  excursion set theory the computation of the halo mass function is mapped
into a first-passage time process in the presence of a  barrier, which
in the spherical collapse model is a constant  and in the ellipsoidal
collapse model is a fixed function of the variance of the smoothed
density field.  However, 
$N$-body simulations show that dark matter halos grow through a
mixture of  smooth accretion,
violent encounters and fragmentations, and modeling halo collapse as spherical, or even as ellipsoidal, is a significant  oversimplification. In addition, the very definition of 
what is a dark matter halo, both in $N$-body simulations and observationally, is a difficult problem.
We propose that some of the physical complications inherent to a realistic description of halo formation can be included in the excursion set theory framework, at least at an effective level, by taking into account that the critical value for collapse is not a fixed constant 
$\d_c$, as in the spherical collapse model, nor a fixed function of the variance $\s$ of the smoothed density field, as in the ellipsoidal collapse model, but rather is itself a stochastic variable, whose scatter reflects a number of complicated aspects 
of the underlying dynamics. Solving the first-passage time problem in the presence of a diffusing barrier we find that the exponential factor in the Press-Schechter mass function changes from $\exp\{-\d_c^2/2\s^2\}$ to $\exp\{-a\d_c^2/2\s^2\}$, where
$a=1/(1+D_B)$ and $D_B$ is the diffusion coefficient of the barrier.  The numerical value of $D_B$, and therefore the corresponding value of $a$, depends among other things on the algorithm used for identifying halos. We discuss the physical origin of the stochasticity of the barrier and, from recent $N$-body simulations that studied the properties of the collapse barrier, we deduce a value $D_B\simeq 0.25$. Our model then predicts  $a\simeq 0.80$, in excellent agreement with the exponential fall off of the mass function found in $N$-body simulations, for the same halo definition.
Combining this result with the non-markovian corrections computed in paper~I of this series, we derive an analytic expression for the halo mass function for gaussian fluctuations and we compare it with $N$-body simulations. 
\end{abstract}



\section{Introduction}

The relation between the
linear  density perturbations at early time and the abundance of virialized 
dark matter halos at the present epoch is an  extremely relevant    one
in modern cosmology. In particular, primordial non-gaussianities leave
an imprint on the abundance and on the clustering properties 
of the most massive objects, such as galaxy clusters, which form out
of rare fluctuations \citep{MLB,GW,LMV,MMLM,KOYAMA,MVJ,RB,RGS}. 
These observational signatures are potentially
detectable by  various planned large-scale galaxy surveys.

From the theoretical side, the challenge is to compute the 
number density of dark matter halos of mass $M$,
$n(M)$, in terms of the statistical properties of the primordial
density field. The formation and evolution of dark matter halos is a highly complicated process, and its full dynamical complexity can only be studied by $N$-body simulations. 
As revealed by $N$-body simulations, halos grow through a
messy mixture of violent encounters, smooth accretion and
fragmentation (see \cite{Springel:2005nw}
and the related movies at 
http://www.mpa-garching.mpg.de/galform/millennium/). 

Still,  some analytic understanding  of halo formation is highly desirable,  both for obtaining a
better physical intuition, and for the flexibility under changes of models or
parameters (such as cosmological model, shape of the non-Gaussianities,
etc.) which is the advantage of analytical results over very timing
consuming numerical simulations. Presently the best available
analytical technique is 
based on Press-Schecther (PS) 
theory \citep{PS} and its extension known as excursion set
theory \citep{Bond} and is able to reproduce, at least qualitatively, several 
properties of dark matter halos seen in $N$-body simulations, such as their conditional and unconditional
mass function,  halo accretion
histories, merger rates and halo bias (see 
\cite{Zentner} for a recent review).
However  excursion set theory describes the collapse as spherical, in its original formulation, or as ellipsoidal,  in the extension due to
\cite{ST}. This is clearly an important oversimplification of the actual complex dynamics and, as a result, while qualitatively the prediction of excursion set theory
agree with $N$-body simulations, at the quantitative level there are important discrepancies, and dynamical evidence in favor of excursion set theory, at least in its present formulation, is quite weak \citep{RKTZ}. A related concern is that numerical simulations show that there is not a good correspondence between peaks in the initial density field and collapsed halos (see \cite{KQG} for an early result).

In this paper we continue the investigation of  excursion set theory that we started  
in \cite{MR1} (hereafter paper~I). In paper~I we have shown how excursion set theory can be put on firmer mathematical grounds, and we have been able to take into account analytically the non-markovian contribution to the evolution of the smoothed density field, due to the use of a tophat filter in coordinate space. In the present paper we turn to a reexamination of the physics behind excursion set theory and we propose a generalization of the theory, based on the idea that the critical value for collapse of the smoothed density field should be treated as a stochastic variable. We discuss  how this stochasticity originates physically and we  show that supplementing excursion set theory with a diffusing barrier allows us to capture at least some of the complexity of the actual halo
formation process, which is lost in the spherical or elliptical collapse model.

Our notation is as in paper~I. Namely, we consider
the density contrast $\d({\bf x}) = [\rho ({\bf x})-\bar{\rho}]/\bar{\rho}$,
where $\bar{\rho}$ is the mean mass density of the universe and ${\bf x}$
is the comoving position,  and  we smooth it
on some scale $R$, defining
\be\label{dfilter}
\d({\bf x},R) =\int d^3x'\,  W(|{\bf x}-{\bf x}'|,R)\, \d({\bf x}')\, ,
\ee
with a filter function 
$W(|{\bf x}-{\bf x}'|,R)$.  For gaussian fluctuations, the statistical
properties of the fundamental density field $\d({\bf x})$ are embodied
in its power spectrum $P(k)$, defined by
\be\label{defP(k)}
\langle \tilde{\d}({\bf k} )\tilde{\d}({\bf k}')\rangle =
(2\pi)^3\d_D({\bf k}+{\bf k}') P(k)\, ,
\ee
where $\tilde{\d}({\bf k} )$ are the Fourier modes of $\d({\bf x})$.
From this one finds the variance $\s^2(R)$ of the smoothed density field
\be\label{mu2RW2}
\s^2(R) \equiv \langle\d^2(R)\rangle\, .
\ee
If we smooth the density field
with a tophat filter function in coordinate
space, the mass $M$ associated to a smoothing radius $R$ is 
$M=(4/3) \pi R^3\rho$,
and we can consider $\s$ as a function of $M$,
rather than of $R$. The ambiguities involved in assigning a mass $M$
to a smoothing scale $R$ when one uses a different filter function
have been discussed in detail in paper~I.

The halo mass function $dn/dM$ can be written as
\be\label{dndMdef}
\frac{dn(M)}{dM} = f(\s) \frac{\bar{\rho}}{M^2} 
\frac{d\ln\s^{-1} (M)}{d\ln M}\, .
\ee
In  Press-Schechter theory \citep{PS} and in excursion set theory
theory \citep{Bond} 
the function $f(\sigma)$ is predicted to be
\be\label{fps}
f_{\rm PS}(\s) = 
\(\frac{2}{\pi}\)^{1/2}\, 
\frac{\d_c}{\s}\, 
\, e^{-\d_c^2/(2\s^2)}\, ,
\ee
where $\d_c\simeq 1.686$ is the critical value in the spherical collapse
model. This result can be extended to arbitrary redshift
$z$ by reabsorbing the evolution of the variance into $\d_c$, so that
$\d_c$ in the above result is replaced by $\d_c(z)=\d_c(0)/D(z)$, where
$D(z)$ is the linear growth factor.
However, \eq{fps} is valid only
if the density is smoothed with a sharp filter in momentum space, and
in this case there is no unambiguous way of assigning a mass to a
region of radius $R$. In paper~I we have been able to extend this
result to a tophat filter in coordinate space. In this case the
computation is considerably more difficult. In fact, when 
the density perturbation is smoothed  with a sharp filter in momentum space,
$\d(R)$ obeys a Langevin equation with respect to the ``pseudotime''
variable $S(R)\equiv \s^2(R)$, with a Dirac delta noise.
This means that 
the dynamics is  markovian, and  that the  probability
$\Pi(\d,S)$ that the density contrast reaches the value $\d$ at
``time'' $S$ satisfies a
Fokker-Planck (FP)
equation, with an ``absorbing barrier'' boundary condition
$\Pi(\d_c,S)=0$.
For different filters the dynamics becomes non-markovian, and 
$\Pi(\d,S)$ no longer satisfies a local diffusion equation such as the
FP equation. In paper~I we have been able to formulate the problem of
the computation of $\Pi(\d,S)$ in terms of a path integral with
boundaries and we have found that the result can be split into a
``markovian'' and a ``non-markovian'' part. The markovian part simply
gives back \eq{fps}, where now $\s^2(M)$ is the variance computed with
the tophat filter in coordinate space, while the non-markovian terms
can be evaluated perturbatively. To first order, we found 
\be\label{ourfI}
f(\s) = (1-\kappa)
\(\frac{2}{\pi}\)^{1/2}\, 
\frac{\d_c}{\s}\, 
\, e^{-\d_c^2/(2\s^2)}
+\frac{\kappa}{\sqrt{2\pi}}\, 
\frac{\d_c}{\s}\, \G\(0,\frac{\d_c^2}{2\s^2}\)\, ,
\ee
where
\be\label{adiRlim}
\kappa (R) \equiv
\lim_{R'\ra \infty}
\frac{\langle\d(R')\d(R)\rangle}{\langle\d^2(R')\rangle} -1
\simeq 0.4592-0.0031\, R\, ,
\ee
$R$ is measured in ${\rm Mpc}/h$, $\G(0,z)$ is the
incomplete Gamma function, and the numerical value
of $\kappa(R)$ is computed assuming
a $\Lambda$CDM model compatible with the WMAP 5yrs data  and
a tophat filter function in coordinate space. Observe that
for a sharp filter in momentum space,
$\langle\d(R')\d(R)\rangle=\langle\d^2({\rm max}(R,R')\rangle$ so
$\kappa(R)$, as defined by the
first equality in \eq{adiRlim}, vanishes.

However, neither \eq{fps} nor \eq{ourfI} perform well when compared to
cosmological $N$-body simulation. 
Indeed,  PS theory predicts too many low-mass halos, roughly by a factor of
two,  and too few high-mass halos: at $\s^{-1}=3$ (high masses
correspond to small values of $\s$), 
PS theory is already
off by  a factor ${\cal O}(10)$. The mass function given in
\eq{ourfI}, in the interesting mass range, is everywhere lower than
the PS prediction 
and therefore, while it improves the agreement at low masses,
it gives an even worse result at high masses, see
Fig.~9 of paper~I. Thus, it is clear that some crucial physical
ingredient is still missing in the model. 
This is not surprising at all, given the use of the simplified spherical collapse model. In the large mass limit the result cannot be improved by turning to the ellipsoidal collapse model since, as we will review in the next section, at large masses the barrier for ellipsoidal collapse reduces to the one for spherical collapse. 
The aim of this paper is to show that  treating the collapse barrier as a stochastic variable allows us to capture some of the complicated physics that is missed by the spherical or ellipsoidal collapse model, and we will see that this modification gives just the required behavior in the large mass limit.

The organization of the paper is as follows. In
Section~\ref{sect:elli}, after  recalling how a moving barrier emerges from the 
ellipsoidal collapse model~\citep{SMT}, we discuss in detail the physical motivations for the introduction of a stochastic barrier.
 In Section~\ref{sect:halo} we
compute the halo mass function with a diffusing barrier, both for the
markovian case and including the non-markovian corrections, and
in Section~\ref{sect:comparison}  we
compare our prediction for the mass function with  $N$-body simulations.
Section~\ref {sect:Concl} contains our conclusions.

\section{The ellipsoidal collapse barrier and the 
diffusing barrier}\label{sect:elli}

The fact that extended PS theory gives a qualitatively correct answer
but fails at the quantitative level
has led many authors either to resort to
fits to the $N$-body simulations, 
see  e.g. \cite{ST,SMT,jenkins,Warren:2005ey,Tinker:2008ff,PPH, grossi2009},  or to look for improvements of the spherical collapse model.
\cite{SMT} took into account the fact that actual halos are triaxial 
\citep {BBKS,BondMyers}, and that the collapse
of halos occurs 
along the principal axes. As a result, the ellipsoidal collapse barrier $B$
acquires a $\s$-dependence, 
\be\label{B(S)}
B(\s)\simeq \delta_c\[ 1 +0.47\(\frac{\sigma}{\delta_c}\)^{1.23}\]\, .
\ee
Physically this reflects the fact  that low-mass halos
(which corresponds to  large $\s$) have larger deviations from
sphericity and significant shear, that opposes  collapse. Therefore
low-mass halos require an higher density to collapse. In contrast,
very large halos are more and more spherical, so their
effective barrier reduces to the one for spherical collapse.

It is apparent that the use of a moving barrier of the form 
(\ref{B(S)}), by itself, cannot improve  the agreement
with $N$-body simulations in the large mass limit since, for large
masses (which correspond to $\s\ra 0$), $B(S)$ reduces to the value
for the spherical collapse and therefore we get back the incorrect
prediction of extended PS theory.  More generally,
since the barrier is receding away from its initial location
$\delta_c$, it is more difficult for the smoothed density perturbation to
reach it, at any $\s$, so the use of \eq{B(S)}  simply gives a halo
mass function which is everywhere smaller than the PS prediction.

In order to improve the agreement between the prediction from the
excursion set method with an ellipsoidal collapse and the $N$-body 
simulations, \cite{SMT} found that
it was necessary to introduce a new parameter $a$ (which, when they
require that their mass function 
fits  the GIF simulation, turns out to have the value
$a\simeq 0.707$,  i.e. $\sqrt{a}\simeq 0.84$) 
and postulate that the form of the barrier is rather
\be\label{B(S)2}
B(\s)\simeq \sqrt{a}\,\, 
\delta_c\[ 1 +0.47\(\frac{\sigma}{ \sqrt{a}\,\delta_c}\)^{1.23}\]\, .
\ee
It is important to stress that, in  \cite{SMT}, 
the parameter  $a$ is not
derived from the dynamics of the ellipsoidal collapse. Rather on the
contrary,  the ellipsoidal collapse model
predicts $a=1$ because in the  limit $\s\ra 0$
the barrier must reduce to that of spherical  collapse.
In \cite{SMT}  the parameter  $a$ is just introduced  by hand 
in order to fit the  $N$-body simulations.

To clarify the origin of this parameter,  it is useful to recall how
\eq {B(S)} emerges. One considers the gravitational collapse of a homogeneous ellipsoid, as in \cite{BondMyers}.  Denoting by $\phi$ 
the peculiar gravitational potential at the location of an ellipsoidal patch,
the deformation tensor 
is $\pa_i\pa_i\phi$,  and
its eigenvalues $\lambda_1,\lambda_2,\lambda_3$ (ordered so that 
$\lambda_1\leq \lambda_2\leq \lambda_3$) characterize the shape of the ellipsoid. In the linear regime, using Poisson equation, the density contrast $\d$ is given by the trace of the deformation tensor, so
$\d=\lambda_1+\lambda_2+\lambda_3$.  In a gaussian random field, the probability distribution of the eigenvalues 
 is known, and is given by (\cite{Dorosk}; see \cite {Lam:2009nd} for a recent generalization to the non-Gaussian case)
 \bees
 p_{\s}(\lambda_1,\lambda_2,\lambda_3)&=&\frac{15^3}{8\pi\sqrt{5}\,\s^6}
 (\lambda_2-\lambda_1) (\lambda_3-\lambda_2) (\lambda_3-\lambda_1)\nn\\
&&\times\exp\left\{-\frac{3I_1^2}{\s^2}+\frac{15I_2}{2\s^2}\right\}
\, ,\label{eq:Dor}
\ees
where $I_1\equiv\ \lambda_1+\lambda_2+\lambda_3=\d$ and
$I_2\equiv\lambda_1\lambda_2+\lambda_2\lambda_3+\lambda_1\lambda_3$. By integrating out $\lambda_1$ and $\lambda_2$ at fixed $\d$, with the constraint $\lambda_1\leq \lambda_2\leq \lambda_3$, one verifies that $\d$ has a gaussian distribution, with variance $\s^2$. 
Rather than using the three eigenvalues as independent variables one can use $\d$, together with the ellipticity $e$ and prolateness $p$, defined by
 \bees
 e &=&\frac{\lambda_3-\lambda_1}{2\d}\, ,\\
 p &=&\frac{\lambda_1+\lambda_3-2\lambda_2}{2\d}\, .
 \ees
From $p_{\s}(\lambda_1,\lambda_2,\lambda_3)d\lambda_1 d\lambda_2 d\lambda_3$ one can derive the distribution probability $g_{\s}(e,p|\d)dedp$ for $e,p$ at fixed $\d$. The result is \citep{BBKS,SMT}
\be
g_{\s}(e,p|\d)=\frac{1125}{\sqrt{10\pi}}\, e(e^2-p^2)
\(\frac{\d}{\s}\)^5\, 
\exp\left\{-\frac{5}{2}\,\frac{\d^2}{\s^2}(3 e^2 +p^2)\right\}\, .
\ee
To define a barrier one needs a criterium for collapse in the ellipsoidal case. In 
\cite{SMT} collapse along each axis is stopped so that the density contrast 
at virialization is the same as in the spherical collapse model, i.e. 179 times the critical density of the universe.  Given this criterium, the critical value for collapse in the ellipsoidal model, $\d_{\rm ell}$, is a function of $e, p$, well approximated by the 
implicit relation \citep{SMT}
\be\label{ST3}
\frac{\d_{\rm ell}(e,p)}{\d_c}\simeq 1+\b\[5(e^2\pm p^2)
\frac{\d^2_{\rm ell}(e,p)}{\d^2_c}\]^{\g}\, ,
\ee
where $\d_c$ is the critical value for spherical collapse, the plus (minus) sign holds for $p$ negative (positive), 
$\b\simeq 0.47$ and $\g\simeq 0.615$. The barrier  (\ref {B(S)}) follows if one replaces $e$ and $p$ with their most probable values according to the distribution $g(e,p|\d)$, which are $\bar{e}=\s/(\d\sqrt{5})$ and
$\bar{p}=0$ (and furthermore one replaces $\d_{\rm ell}(e,p)$, on the right-hand side of \eq{ST3}, with $\d$).

As we already mentioned, this barrier always stays above the spherical collapse barrier at $\d_c$,  and reduces to the spherical collapse one for $\s=0$, i.e. for large masses. 
For our purpose, it is however important to note that this result only
holds if $e$ and $p$ are replaced by their most probable values
$\bar{e}$ and $\bar{p}$. For generic values of $e$ and $p$  the
critical value for collapse can be either higher or {\em lower} than
$\d_c$. This results in a "fuzzy" threshold \citep{Audit,LeeS,SMT},
with a probability distribution that extends 
even to values smaller than $\d_c$. To compute the variance of the barrier  due to this effect we use a slightly more accurate expression for the critical value of the ellipsoidal collapse as a function of the 
eigenvalues $\lambda_i$ \citep{Sandvik},
\be\label{Blambda}
\d\sub{ell}(\lambda_1,\lambda_2,\lambda_3)=\d_c\[ 1+\a_1 (\lambda_2-\lambda_1)^{\a_2}+
\a_3 (2\lambda_3-\lambda_2-\lambda_1)^{\a_4}\]\, ,
\ee
where $\a_1\simeq 0.2809$, $\a_2=1.3557$, $\a_3\simeq 0.070$, $\a_4\simeq 1.41205$.\footnote{As in \eq {B(S)}, we are considering the barrier at $z=0$, when the linear growth factor $D(z)=1$. The expression for generic $z$ is obtained by rescaling
$\d_c\ra \d_c/D(z)$ and $\lambda_i\ra\lambda_i D(z)$, see 
eq.~(18) of \cite{Sandvik}.} We stress that this expression for the critical value of $\d$   was found in
\cite {Sandvik} requiring an accurate representation of the ellipsoidal collapse of
\cite{BondMyers}, and not by fitting to $N$-body mass functions.
The average value of this barrier is 
\bees
&&B\sub{ell}(\s)\equiv \langle \d\sub{ell}(\lambda_1,\lambda_2,\lambda_3)\rangle \label{Bave}\\
&=&\int_{-\infty}^{\infty}d\lambda_3
\int_{-\infty}^{\lambda_3}d\lambda_2\int_{-\infty}^{\lambda_2}d\lambda_1
\, \d\sub{ell}
(\lambda_1,\lambda_2,\lambda_3)p_{\s}(\lambda_1,\lambda_2,\lambda_3)
\, ,\nn
\ees
where  $p_{\s}(\lambda_1,\lambda_2,\lambda_3) $ is the 
probability distribution given in  \eq{eq:Dor}. 
In Fig.~\ref {fig:diff_barrier_ave}  we compare this expression for 
$B\sub{ell}(\s)$ with the expression given in ({\ref{B(S)}). We see that they provide two similar representation of the average barrier for ellipsoidal collapse.
The variance of $B\sub{ell}(\s)$ is given by

\begin{figure}
\includegraphics[width=0.4\textwidth]{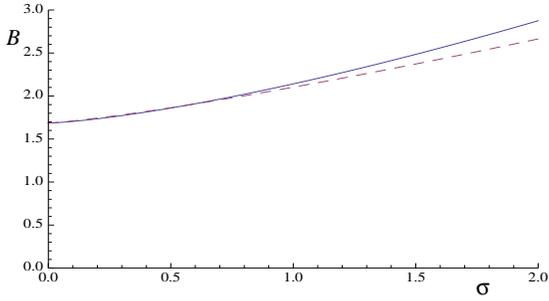}
\caption{\label{fig:diff_barrier_ave}
Two possible representations for the 
ellipsoidal collapse barrier  as a function of $\s$.
The red dashed line
is the barrier given in \eq{B(S)}. The blue solid line is the function $B\sub{ell}(\s)$ obtained from \eqs{Blambda}{Bave}.
}
\end{figure}

\bees
\Sigma^2_{B\sub{ell}}(\s)&\equiv& \langle 
[\d\sub{ell}(\lambda_1,\lambda_2,\lambda_3)-B\sub{ell}(\s)]^2\rangle
\label{SigmaBlambda}\\
&\simeq & \d^2_c  \[
0.00805 \s^{2 \a_2} + 0.00489 \s^{2 \a_4} + 
0.00305 \s^{\a_2 + \a_4}\]\nn\, ,
\ees
where again the averages have been computed 
using the distribution (\ref {eq:Dor}). In Fig.~\ref {fig:diff_barrier_3Sigma} we compare $B\sub{ell}(\s)$ with the curves
$B\sub{ell}(\s) \pm 3\Sigma_{B\sub{ell}}(\s)$. We see that a fluctuation of the barrier at the $3\Sigma_{B\sub{ell}}$ level can bring the threshold for collapse well below the constant value $\d_c$ derived from the spherical collapse model (see also Fig.~7 of \cite{SMT}).

\begin{figure}
\includegraphics[width=0.4\textwidth]{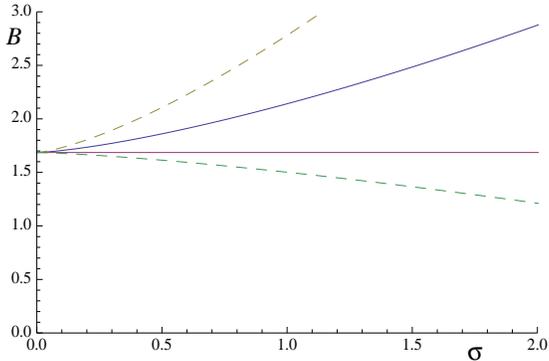}
\caption{\label{fig:diff_barrier_3Sigma}
The  function $B\sub{ell}(\s)$ (blue solid line), together with  the curves
$B\sub{ell}(\s) \pm 3\Sigma_{B\sub{ell}}(\s)$ (dashed lines). The  horizontal red solid line is the spherical collapse barrier $B=\d_c$.
}
\end{figure}

This result already makes it clear that, as a matter of principle, the critical value for collapse is unavoidably a stochastic variable. However,  the fluctuations of the barrier 
discussed above by no means exhaust all possible sources of stochasticity in the actual physical problem.  For instance, halos are  subject to tidal effects due to their environment, which also results in a distribution of values for the collapse barrier
\citep{Desjacques}. More generally,
modeling  dark matter halos as  smooth and homogeneous ellipsoids characterized by the eigenvalues $\lambda_i$, even  when taking into account  their distribution probability, 
is still a significant  oversimplification.  
For instance, a patch that is collapsing  might have significant non-linear substructures, whose presence influences its critical value for collapse.
All these effects contribute to the scatter of the values of the threshold for collapse. 

Last but not least, the very definition of what is a dark matter halo
is a non-trivial problem both in
numerical simulations and  observationally
(for cluster observations, see \cite{Jeltema} and references therein).
In simulations, halos are  usually identified either through  a 
Friends-of-Friends (FOF) algorithm, or using spherical overdensity 
(SO) finders. 
However actual halos are triaxial, rather than spherical, and often
messier than that, and there is nothing fundamental or rigorous  in
either choice, both being largely a matter of convenience. FOF halo
finders track isodensity profiles and might be more relevant for
Sunayev-Zeldovich or weak lensing, while SO finders may be more
relevant for cluster work. 
Searching  for
halos using for instance a spherical overdensity  finder, when 
halos are at best triaxial and often more irregular, 
introduces a further source 
of statistical fluctuations, both in the number count of halo, 
and in the assignment of the mass. A similar concern is that the exact
definition of a virialized halo depends on the what one means exactly 
by ``virialized''.

So in the end, in a given $N$-body
simulation, each patch of the initial density field that eventually
collapses to form a halo at a given epoch, has a smoothed overdensity
that  does not have in general exactly the value predicted by the ellipsoidal
collapse model, but rather fluctuates around it with fluctuations
that are determined by various factors, such as the distributions of
the eigenvalues of the deformation tensor,  
the details of the halo finder algorithm, or other details related to the environment, the presence of non-linear substructures, etc., as discussed above.

Motivated by these considerations, we propose in this paper  to extend excursion set theory by considering a first-passage time problem in the presence of a  barrier that fluctuates stochastically. 
Given that the fluctuations in the collapse
threshold depend, among other things, on the exact details of the halo
definition (halo finder, virialization critierium, etc.), the mass
function computed from excursion set theory with such a stochastic barrier will
depend on these details, too. This is a positive aspect because 
the actual halo mass function obtained from $N$-body simulations 
depends on the halo finder~\citep{White}. For instance, with FOF finders
the mass function depends on the link-length used, and in
particular  the value $a\simeq 0.707$ given in \eq{B(S)2}
holds for a link-length equal to 0.2 times the mean inter-particle 
separation \citep{SMT}.\footnote{Observe that, for this link length, a sizable 
fraction of halos have major non-spherical substructures and a 
significant contribution to the halo mass arises from outside the
"virial" radius \citep{Lukic:2008ds}. This underlines again the
importance of the details of the definition of what is a halo.}
When halos are identified with a SO finder, the mass function 
depends on the value chosen for the overdensity $\D$, 
see e.g.~\cite {Tinker:2008ff}. These effects cannot be reproduced by
excursion set theory with a barrier which is fixed uniquely by the dynamics of
the spherical or ellipsoidal collapse, and which therefore is
insensitive to these details. In this paper we explore whether the use of a
stochastic barrier allows us to incorporate into excursion set 
theory,  at least at the level of an
effective description, a part of the
stochasticity intrinsic to the actual physical problem of halo
formation, and due both to the complicated underlying dynamics and to
the choices that one has made when giving an operative definiton of
dark matter halos.

Ideally, one would like to compute theoretically 
the fluctuation properties of the barrier. For some effect, such as
that due to the distribution of eigenvalues of the deformation
tensor, this is possible, as we
saw in \eq{SigmaBlambda}. Unfortunately, other effects such as 
the scatter of the barrier 
due to the details of the halo finder (which, as it will turn out, 
give a
contribution that dominates over that in \eq{SigmaBlambda}) 
are much more difficult to predict theoretically. 
We will therefore take a more phenomenological approach. We will
consider a barrier that performs a random walk with a diffusion
coefficient $D_B$. At least at a first level of description, all our
ignorance of the details of the dynamics of halo formation is buried
into this coefficient. Solving the first-passage time problem with
such a barrier we will find that the net effect is that
in the halo mass function 
predicted by Press-Schecther or excursion set theory 
the exponential factor
changes from
$\exp\{-\d_c^2/(2\s^2)\}$ to $\exp\{-a\d_c^2/(2\s^2)\}$
(and more generally we must replace
everywhere $\d_c\ra\d_c\sqrt{a}$), where 
\be\label{apred}
a=\frac{1}{1+D_B}\, ,
\ee 
see Section~\ref{sect:halo}. This is just the replacement that was postulated in \cite{SMT,ST} in order to fit the data. We therefore discover that the Sheth and Tormen (ST) mass function (at least in the large mass limit), is just the mass function obtained by excursion set theory with a diffusing barrier.

Having obtained a physical understanding of the parameter $a$ that appears in the ST mass function, one can ask whether it is possible to go beyond the approach in
\cite{ST,SMT}, where $a$ is simply treated as a fitting parameter, and try to predict it, by computing the diffusion coefficient $D_B$.
Given that   a first-principle computation of $D_B$ seems difficult,  
in Section~\ref{sect:comparison} we
will turn to $N$-body simulations themselves. We will see that
recent numerical studies of the properties of the collapse threshold allows us
to deduce the value of $D_B$. Given this input we will then  
compare our prediction (\ref{apred})
to the slope of the exponential fall of of the mass function, and more
generally (including in the
halo mass function also the non-markovian corrections computed in
paper~I) we will compare our analytic form for the mass function to 
the numerical data.  Even if in this way we must use an input from  $N$-body
simulations themselves, still the comparison is quite non-trivial,
since the relation (\ref{apred}) is a specific prediction of our model.

\section{The halo mass function in the presence of the diffusing barrier}
\label{sect:halo}

In order to illustrate the idea in a simple mathematical setting, we consider a barrier that fluctuates over the constant spherical collapse barrier  $B=\d_c$. In the large mass limit the ellipsoidal collapse barrier reduces to the spherical collapse barrier, so we expect that this approximation should be adequate for computing the effect of a stochastic barrier on the 
high-mass tail of the halo mass function.  We also consider a 
barrier $B$  that fluctuates in such a way that its mean root square fluctuation $\Sigma_B$ depends linearly on the variance of the smoothed density field $\s(R)$,
\be\label{SigmaBDB}
\Sigma_B(\s(R))\equiv \langle\left(B-\langle B\rangle\right)^2\rangle^{1/2}
=\s(R)\sqrt{D_B}\, ,
\ee
where $D_B$ is a numerical coefficient. This choice is partly motivated by mathematical simplicity. Furthermore, we will see in Section~\ref{sect:comparison} that there is some evidence from $N$-body simulations that for small $\s(R)$ the barrier diffuses just as in \eq {SigmaBDB}.

This form of the barrier corresponds to a brownian motion.
Recall in fact that,
if a particle performs a one-dimensional brownian motion, its position $x(t)$
has a  variance given by 
$\langle(x(t)-x_0)^2\rangle^{1/2}=\sqrt{D t}$, where $D$ is the diffusion
coefficient.
In the excursion set method  $S(R)=\s^2(R)$ plays
the role of a ``pseudotime'' variable, so \eq {SigmaBDB} means that the collapse barrier
performs a brownian motion around its initial position $\langle B\rangle$,
with a diffusion coefficient $D_B$.  We  will refer to
the model in which the barrier's scatter behaves as in \eq{SigmaBDB} as a
 "diffusing barrier". As we will see below,  the first passage time problem in the presence of a diffusing barrier can be solved analytically, so 
 \eq {SigmaBDB} provides at least a useful toy model for understanding the effect of a more general stochastic barrier.  We emphasize however that  the idea that we are proposing is more general, and can in principle we implemented for a generic functional form of  the barrier $B(\s)$ and of its fluctuation $\Sigma_B(\s)$,  although the associated first-passage time problem becomes more complicated, and might require the  numerical generation of an ensemble of trajectories using  Monte Carlo simulations,
 as in \cite{Bond}.

Intuitively, we can  understand why a diffusing barrier
can help to reproduce the numerical $N$-body results. 
We are interested in events, corresponding to
cluster masses, that arise from rare fluctuations, on the far tails of the
probability distribution. For instance at $\sigma^{-1}=3$, the PS theory
prediction for $f(\sigma)$ is about $10^{-5}$, and we are searching for
a mechanism that brings this number up to the observed value of about 
$10^{-4}$. 
Even if,
on average, a barrier has equal probabilities of fluctuating toward
values lower that $\d_c$ as toward values higher than $\d_c$,
still the fluctuations of the barrier toward lower values can
have a much more significant effect (consider for instance
what happens to a dam if on a rare occasion it is lowered).
In fact, this is true even if most of the flucutation where above
$\d_c$, and only rare flucutations were below $\d_c$ which, as we
will see, is the case when we consider fluctuations over the
ellipsoidal collapse barrier, see also
Fig.~\ref{fig:diff_barrier_3Sigma}. In the analogy with the dam, rare
occasional lowerings of the dam can produce substantial flooding, even if
many more flucutations rather raise it.

To verify formally  this intuition,  we neglect at first the non-markovian corrections due to the filter, discussed in paper~I. We denote by
$\Pi (B(0),B;\d(0),\d;S)$ the  joint probability that, at ``time'' $S$,
the barrier has reached by diffusion the value $B$, starting from the 
initial value $B(0)=\d_c$, while
the density contrast has reached the value $\d$, starting from the 
initial value $\d(0)=0$.  In the markovian case the probability distribution obeys a Fokker-Planck equation. The fact that the
``particle'' described by $\d(S)$ and the barrier $B(S)$ both diffuse
independently means that the joint probability distribution satisfies
the two-dimensional FP equation
\be\label{FPdS2dim}
\frac{\pa\Pi}{\pa S}=\frac{D}{2}\, \frac{\pa^2\Pi}{\pa \d^2}
+\frac{D_B}{2}\, \frac{\pa^2\Pi}{\pa B^2}
\, .
\ee
In our case the diffusion coefficient of  $D=1$, see e.g. eq.~(20) of paper~I, while
 $D_B$ is the diffusion coefficient of the barrier. To solve this equation it is convenient  to introduce a  "time" variable $t=S/\d_c^2$, and
the variables
\bees
x_1&=&\frac{\d_c-B}{\d_c\sqrt{D_B}}\, ,\\
x_2&=&\frac{\d_c-\d}{\d_c}\, ,
\ees
so \eq{FPdS2dim} becomes
\be\label{FPdS2dimy}
\frac{\pa\Pi}{\pa t}=\frac{1}{2}\, \frac{\pa^2\Pi}{\pa x_1^2}
+\frac{1}{2}\, \frac{\pa^2\Pi}{\pa x_2^2}
\, .
\ee
In term of these variables the barrier starts at $x_1(0)=0$ while the
``particle'' starts at $x_2(0)=1$. The
boundary condition is that
$\Pi (B(0),B;\d(0),\d;S)$ vanishes when 
$\d(S)=B(S)$, i.e. $\Pi (x_1(0),x_1;x_2(0),x_2;t)$ vanishes when
$\sqrt{D_B}\,x_1=x_2$. We define $\theta$ from $\sqrt{D_B}=\tan\theta$, so we
have a two-dimensional FP equation with the
boundary condition that $\Pi$ vanishes on the line $x_1=x_2\cot\theta$,
see Fig.~\ref{image}.
This problem can be solved by the
method of images  \citep{redner2001}, and the result
is given by a gaussian centered on
$(x_1=1,x_2=0)$ minus a gaussian centered on the image point
$(x_1=\sin 2\theta,x_2=-\cos 2\theta)$, 
\begin{eqnarray}
&&\Pi^{\rm gm}(x_1(0)=0,x_1;x_2(0)=1,x_2;t)\nn\\
&&=\frac{1}{2\pi t}
\left[e^{-[x_1^2+\left(x_2-1\right)^2]/2 t}-
 e^{-[\left(x_1-\sin 2 \theta\right)^2
+\left(x_2+\cos 2 \theta\right)^2]/2 t}\right]\, ,
\end{eqnarray}
where, as in paper~I, we added to $\Pi$ the superscript ``gm'' to remind that this
is the solution for gaussian fluctuation with a  markovian dependence on the smoothing scale

\begin{figure}
\includegraphics[width=0.4\textwidth]{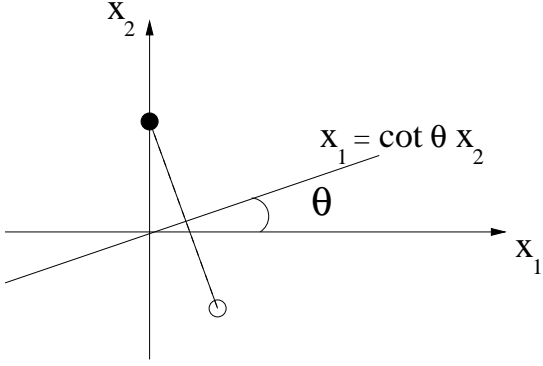}
\caption{\label{image}
Mapping of the scaled density perturbation and collapse barrier
coordinates  in one dimension to the 
plane $(x_1,x_2)$. The initial position is in $(x_1=0,x_2=1)$ (black
dot) and its image point is in $(x_1=\sin 2\theta,x_2=-\cos 2\theta)$
(white dot).
}
\end{figure}

The probability density for the scaled density perturbation to be 
at the position $x_2$ is the integral of the two-dimensional
density over the accessible range of the scaled collapse barrier 
coordinate $x_1$,
\begin{eqnarray}
&&\Pi^{\rm gm}(x_2,t)
=\int_{-\infty}^{x_2\,\cot\theta}\, dx_1\, 
\Pi^{\rm gm}(x_1;x_2;t)\nonumber\\
&=&\frac{1}{2\sqrt{2\pi t}}\left[e^{-(x_2-1)^2/2 t}\,{\rm Erfc}
\left(-\frac{x_2\,\cot\theta}{\sqrt{2\,t}}\right)\right.\nonumber\\
&&-\left. e^{-(x_2+\cos 2 \theta)^2/2 t}\,{\rm Erfc}
\left(\frac{\sin 2\theta -x_2\,\cot\theta}{\sqrt{2\,t}}\right)
\right]\, , 
\end{eqnarray}
where ${\rm Erfc}(z)$ is the complementary error function and the
initial conditions $x_1(0)=0$ and $x_2(0)=1$   are understood.
Restoring the original
variables $S,\d(S)$ and $B(S)$, and
using $\Pi(\d_0;\d;S)d\d=\Pi(x_2(0);x_2;t) |dx_2|$ where
$|dx_2|=d\d/\d_c$, we get 
\bees
&&\Pi^{\rm gm}(\d,S)=\frac{1}{2\sqrt{2\pi S}}\, \left\{
e^{-\d^2/(2S)} 
{\rm Erfc}\[-\frac{\cot\theta}{\sqrt{2S}} (\d_c-\d)\]\right.\nn \\
&&\left. 
-e^{-(2\d_c\cos^2\theta-\d)^2/(2S)}
{\rm Erfc}\[\frac{\d_c\sin2\theta-(\d_c-\d)\cot\theta}{\sqrt{2S}} \]
\right\}\, ,\label{Pimarkovth}
\ees
where the initial condition $\d_0=0$ is understood. The limit of
non-diffusing barrier is $D_B\ra 0^+$, so $\theta\ra 0^+$
and $\cot\theta\ra +\infty$. Recalling
that ${\rm Erfc}(z)\ra 2$ as $z\ra -\infty$, we see that in the limit
$D_B\ra 0^+$ we recover the standard result of excursion set theory
with a static barrier of \cite{Bond}.

\begin{figure}
\includegraphics[width=0.4\textwidth]{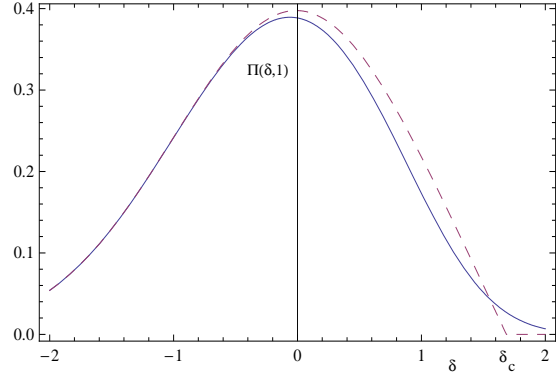}
\caption{\label{fig:Pidiffusing}
The function $\Pi^{\rm gm}(\d,S)$ with  $D_B=0.25$ (blue
solid line) compared to the standard excursion set result, i.e.
$\Pi^{\rm gm}(\d,S)$ with
$D_B=0$ (violet dashed line), as functions of $\d$, for $S=1$.
}
\end{figure}

In Fig.~\ref{fig:Pidiffusing} 
we compare this function, for a diffusion coefficient
$D_B=0.25$, with the
static barrier case. Observe that, when $D_B=0$, the distribution
function vanishes for $\d\geq \d_c$,\footnote{This only holds because,
for the markovian term, we can work directly in the continuum limit.
As we discussed in
paper~I, if we compute $\Pi$ summing over trajectories defined
discretizing time in steps $\eps$, there are finite $\eps$ corrections
and $\Pi(\d,S)$ non longer vanishes for $\d\geq \d_c$, even  for the
static barrier.} while for finite $D_B$ it is non-zero for all values
of $\d$. Of course, this reflects the fact that the barrier can
in principle diffuse to arbitrarily large values of $\d$.

The markovian contribution to the first crossing
rate is 
\be\label{defcalF}
{\cal F}^{\rm gm}(S) 
= -\int_{-\infty}^{\infty}\,d\d\, 
\frac{\pa \Pi^{\rm gm}(\d,S)}{\pa S}\, .
\ee
The evaluation of this expression 
can be simplified observing that $\pa/\pa S$, when acting on
$(2\pi S)^{-1/2} \exp\{-\d^2/(2S)\}$, is the same as $(1/2)\pa^2/\pa\d^2$,
and integrating twice by parts $\pa^2/\pa\d^2$. We then find
\be
{\cal F}^{\rm gm}(S)=
\frac{\delta_c}{\sqrt{2\pi(1+D_B)}\, \,S^{3/2}}\,
\exp\left\{-\frac{\delta_c^2}{2(1+D_B)S}\right\}\, .
\ee
The function $f(\sigma)$ is obtained from the first crossing rate 
using $f(\sigma)=2\sigma^2{\cal F}(\sigma^2)$, see e.g. Section~2 of
paper~I, so we get
\be\label{fmarkov}
f^{\rm gm}(\s) = 
\(\frac{2}{\pi}\)^{1/2}\, 
\frac{\sqrt{a}\, \d_c}{\s}\, 
\, e^{-a\d_c^2/(2\s^2)}\, ,
\ee
where
\be\label{a79}
a=\frac{1}{1+D_B}\, .
\ee
This is the crucial result of this section. We see that the effect of the diffusing barrier is that the exponential factor in the halo mass function changes from
$\exp\{-\d_c^2/(2\s^2)\}$ to $\exp\{-a\d_c^2/(2\s^2)\}$, with $a$ given by 
\eq{a79}, and more generally everywhere in the mass function $\d_c\ra\d_c\sqrt{a}$. This is exactly the modification which was postulated {\em ad hoc} in \cite{SMT}.

In fact, even if the expression for
$\Pi^{\rm gm}(\d,S)$ given in \eq{Pimarkovth} is interesting by itself,
the result for the first-crossing rate
could have been obtained directly, without
even computing explicitly 
$\Pi^{\rm gm}(\d,S)$, simply by observing that the problem involving
a barrier with coordinate $x_1$ and diffusing with a 
diffusion coefficient $D_1$, and a particle
with coordinate $x_2$, diffusing with a 
diffusion coefficient $D_2$,
can be mapped into a one-degree of freedom problem,  introducing the
relative coordinate $x=x_2-x_1$. The resulting stochastic motion
is governed by an effective diffusion coefficient $D_{\rm eff}=D_1+D_2$
\citep{redner2001}.  This point can be easily
understood considering a Langevin equation for the barrier coordinate
$x_1$, 
\be
\dot{x}_1=\eta_1(t)\, ,
\ee
with
\be
\langle \eta_1(t)\eta_1(t')\rangle =D_1\, \d(t-t')\, ,
\ee
and a Langevin equation for the particle coordinate
$\dot{x}_2=\eta_2(t)$ 
with $\langle \eta_2(t)\eta_2(t')\rangle =D_2\, \d(t-t')$. Then the
relative coordinate $x=x_2-x_1$ satisfies 
$\dot{x}=\eta(t)$ with $\eta(t)=\eta_2(t)-\eta_1(t)$ and, if 
$\eta_1(t)$ and $\eta_2(t)$ are uncorrelated,
\bees
\langle \eta(t)\eta(t')\rangle &=&
\langle \eta_1(t)\eta_1(t')\rangle +
\langle \eta_2(t)\eta_2(t')\rangle\nn\\
&=&(D_1+D_2)\, \d(t-t')\, ,
\ees
showing that the relative coordinate diffuses with an effective
diffusion coefficient $D_1+D_2$. In our
case $D_1=D_B$ and $D_2=1$.

We have repeated the above analysis including the non-markovian corrections 
due to a tophat filter in coordinate space,
to first order, using the results obtained in paper~I. The explicit computation is performed in Appendix~A. The result is
\be\label{ourf}
f(\s) = (1-\tilde{\kappa})
\(\frac{2a}{\pi}\)^{1/2}\, 
\frac{\d_c}{\s}\, 
\, e^{-a\d_c^2/(2\s^2)}
+\frac{\tilde{\kappa}\d_c\sqrt{a}\,}{\s\sqrt{2\pi}}\,
 \G\(0,\frac{a \d_c^2}{2\s^2}\)\, ,
\ee
where
\be\label{tildekappa}
\tilde{\kappa}=\frac{\kappa}{1+D_B}\, .
\ee
This is our result for the halo mass function. We hasten to add that this result only holds in the large mass limit, otherwise we must consider fluctuations over the ellipsoidal barrier, rather than over the constant spherical constant barrier, and furthermore we have considered a specific form of the barrier variance, corresponding to a random walk. We now turn to a comparison of our model with $N$-body simulations.

\section{Comparison with $N$-body simulations}\label{sect:comparison}

The variance of the collapse barrier in $N$-body simulations has been recently studied in \cite{RKTZ}
(see also\cite{Dalal:2008zd}). 
For each halo identified in their $N$-body simulations at $z=0$, they
calculated the center-of-mass of the halo particles at 
their positions in the linear field at $z\simeq 10^2$ and used the
density field, smoothed  with  a tophat filter in real space,
to compute the overdensity within a given lagrangian radius $R$.
This overdensity is then linearly extrapolated  to $z=0$. They find
that the
distribution of such smoothed
linear overdensities $B(\s)$,  at fixed $\sigma$,
is approximately log-normal in shape with a width $\Delta_{B}$ given
by
\be\label{SigmaB}
\D_{B}\simeq 0.3\,\sigma\, .
\ee
In a log-normal distribution one has
\be\label{var1}
\Sigma_B\equiv\langle\left(B-\langle B\rangle\right)^2\rangle^{1/2}=\left(e^{\D^2_{B} }-1
\right)^{1/2}\,\langle B \rangle\, ,
\ee
and in our case, for small $\s$, 
$\langle B\rangle\simeq \delta_c\simeq  1.68$.
Observe that \eq{var1} is consistent with our estimate (\ref{SigmaBlambda}). In fact,   \eq{SigmaBlambda} only includes the scatter due to the fluctuations of the eigenvalues, so it is a lower bound on the actual scatter of the barrier that, as we discussed in Section~\ref{sect:elli},  can  in principle receive contributions from many other effects. As we see from Fig.~\ref{fig:variances} the variance given by \eq{var1} is indeed always above that given by \eq{SigmaBlambda}. 

In a $\Lambda$CDM model, $\s(M)$ is such that, for values of $M$
corresponding to cluster of galaxies, $0.3\s(M)\ll 1$. For instance,
$\s(M)=1$ for $M\simeq 10^{14}\msun h^{-1}$, while
$\s(M)=0.6$ for $M\simeq 10^{15}\msun h^{-1}$ (see e.g. Fig.~1 of the review 
\cite{Zentner}). Therefore in the high-mass range $\D_B$ is small
and we can expand \eq{var1}, obtaining
\be\label{var2}
\Sigma_B\simeq
 \langle B\rangle \D_{B}\simeq 0.3\,\delta_c\,\sigma\, .
\ee 
The inclusion of an overall drift of the barrier, such as that in
\eq{B(S)}, as well as higher order terms in the expansion of the
exponential in \eq{var1},
provides terms of higher order in $\s$, which are
subleading in the large-mass regime. 

\begin{figure}
\includegraphics[width=0.4\textwidth]{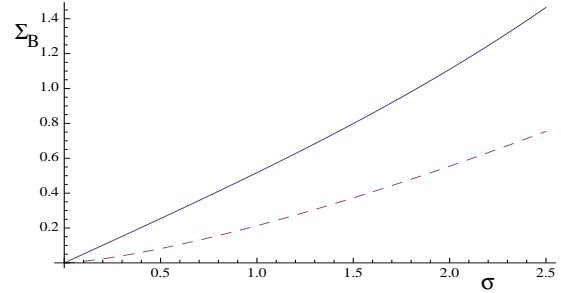}
\caption{\label{fig:variances}
The variance of the barrier $\Sigma_B$ from \eq{var1}  (blue solid line)
compared to the estimate (\ref {SigmaBlambda}) (violet dashed line).
}
\end{figure}

\Eq {var2} has exactly the form of the diffusing barrier given in 
eq.~(\ref {SigmaBDB}), with
a diffusion coefficient 
\be\label{Dc025}
D_B\simeq (0.3\,\delta_c)^2\simeq 0.25\, .
\ee
In this case our model predicts a relation, given by \eq {a79}, between the diffusion coefficient $D_B$ of the barrier and the slope of the exponential of the halo mass function in large mass limit,  i.e. we predict
\be\label{a80}
a=\frac{1}{1+D_B}\simeq 0.80\, .
\ee
The value (\ref {Dc025}) has been deduced from the $N$-body simulations of
\cite{RKTZ}, where  halos where identified with a $\D=200$ spherical overdensity algorithm. We therefore must compare our prediction with the value of $a$ obtained under the same conditions. This can be obtained from \cite{Tinker:2008ff}, where the same numerical simulation was used to study the halo mass function (and its deviations from universality, see below). The authors fit their result with a fitting function
\be\label{fitfs}
f(\s) = A_T\[ \(\frac{\s}{b_T}\)^{-a_T}+1\] e^{-c_T/\s^2}
\ee
and, for a spherical overdensity $\D=200$, they find $A_T=0.186$,
$a_T=1.47$, $b_T=2.57$ and $c_T\simeq 1.19$. We add a subscript $T$,
which stands for Tinker et al., to distinguish for instance their parameter $a_T$ from our parameter $a$.
A first indication of the agreement of our prediction with the above fitting formula can be obtained by comparing the respective exponential cutoff.  
In terms of their parameter $c_T$,  our parameter 
$a$ is given by the combination $2c_T/\d_c^2$. Using their value
$c_T\simeq 1.19$, one has
${2c_T}/{\d_c^2}\simeq 0.837$, in  good agreement with the value (\ref{a80}).  
This agreement is a non-trivial result.  It is true that,
to get  \eq{a80}, we used an input from the same $N$-body simulation,
namely the 
scatter of the values of the threshold for collapse, from which we
deduced the
diffusion coefficient $D_B$ of the  barrier. However, 
given this input our model makes a   non-trivial prediction for the
the numerical value of the parameter $a$ that appears in the halo mass
function. This is very different from fitting $a$ directly to the
$N$-body simulations. In principle,
the prediction $a=1/(1+D_B)$ could have given rise to a  value of $a$ 
very different from the one extracted  directly from 
$N$-body simulations, and this would have falsified the diffusing barrier model.

Of course, given that the functional forms of $f(\s)$ in
\eqs{ourf}{fitfs} are different (which also implies that the normalization constant $A_T$ in \eq{fitfs} is not exactly the same as the overall factor that we have in front of the exponential in \eq{ourf}), 
the proper way of performing an accurate comparison is not in terms of the location of the exponential cutoff, but rather directly in terms of the full functions $f(\s)$.
In Fig.~\ref {fig:compare_Nbody} we compare our prediction 
for $f(\s)$, given by \eq{ourf}, to the  function
(\ref{fitfs}) representing the fit to the $N$-body simulation. 
Observe that the vertical axis ranges over more than three orders of magnitudes.

\begin{figure}
\includegraphics[width=0.4\textwidth]{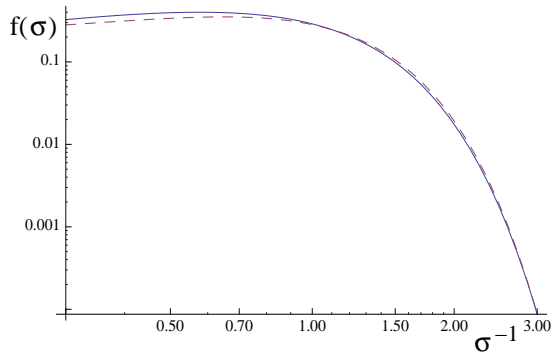}
\caption{\label{fig:compare_Nbody}
Our prediction for $f(\s)$ given in \eq{ourf} (blue solid line) compared to the fit
to $N$-body simulations given by \eq{fitfs} (violet dashed line), in a log-log scale.
}
\end{figure}

\begin{figure}
\includegraphics[width=0.4\textwidth]{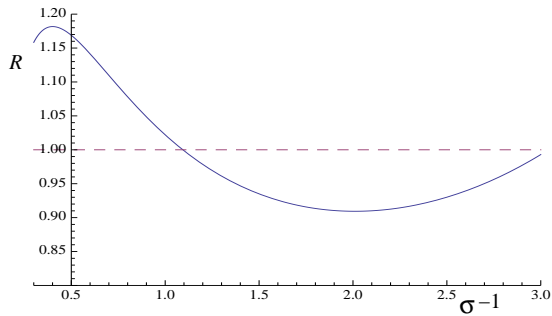}
\caption{\label{fig:compare_Nbody_ratio}The ratio $R$ between our prediction 
(\ref{ourf}) for 
$f(\s)$ and  the fit
to $N$-body simulations given by \eq{fitfs} (blue solid line). The dashed line 
marks the line $R=1$.
}
\end{figure}

To make a more detailed comparison in Fig.~\ref{fig:compare_Nbody_ratio} we plot, on a linear-linear scale, the ratio $R$ between our prediction for 
$f(\s)$ and the Tinker et al. fit
to $N$-body simulations given in \eq{fitfs}.  We see that for all values of $\s^{-1}\geq 0.3$ the discrepancy between our analytic result and the fit to  the $
N$-body simulation is 
smaller than  $20\%$, and for $\s^{-1}\geq 1$ it is smaller than
$10\%$. Considering that our result comes from an analytic model of halo formation with no tunable parameter (the parameter $a$ is fixed once $D_B$ is given, and we do not have the right to tune it),  while \eq {fitfs} is  simply a fit to the data with four free parameters, we think that this result is quite encouraging. The numerical accuracy
is actually the best that one could have hoped for, considering for instance that we have neglected  second-order non-markovian corrections. From
\eq{ourf} we see that, in the computation of the non-markovian effect due to the  tophat filter in coordinate space  in the presence of a diffusing barrier,  the actual expansion parameter is $\tilde{\kappa}=\kappa/(1+D_B)$ which, using \eq {adiRlim} with $R=10$~Mpc and $D_B\simeq 0.25$, has a numerical value $\tilde{\kappa}\simeq 0.34$. Therefore the second-order non-markovian corrections,
which are proportional to $\tilde{\kappa}^2$, are expected to be of order
$10\%$. Furthermore, as  we move toward lower masses the effect of the ellipsoidal barrier must become important,  while \eq{ourf} has been obtained using the spherical collapse model, and the variance of the barrier $\Sigma_B$ shown in 
Fig.~(\ref {fig:variances}) (solid line) has been approximated by a straight line
in \eq {var2}, which again is only valid at small $\s$.

\begin{figure}
\includegraphics[width=0.4\textwidth]{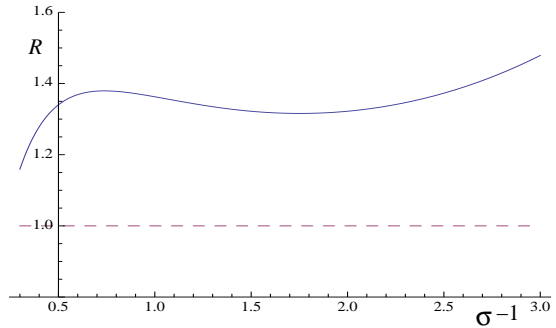}
\caption{\label{fig:compare_Nbody_nokappa}The ratio $R$ between our prediction 
(\ref{ourf}) for 
$f(\s)$ and  the fit
to $N$-body simulations given by \eq{fitfs} (blue solid line), setting $\kappa=0$. The dashed line 
marks the line $R=1$.
}
\end{figure}

In Fig.~\ref{fig:compare_Nbody_nokappa}  we show the result that one obtains for the ratio $R$ if one includes the diffusing barrier but neglects the non-markovian corrections due to the tophat filter in coordinate space, i.e. if one sets $\kappa =0$ in \eq{ourf}.  We see that the agreement degrades, and the discrepancy  becomes of order 30-50\%. Thus, while the largest part of the improvement, compared to 
PS theory, comes from the introduction of a diffusing barrier (recall that PS theory, which predicts $a=1$, is off by one order of magnitude in the high-mass limit, see e.g. Fig.~1 of paper~I), still
for an accurate computation it is important to include the non-markovian corrections due to  the tophat filter in coordinate space. Observe also that,
in the large mass limit, the term proportional to the 
incomplete Gamma function in \eq{ourf} is subleading, and the effect of the filter
is basically to reduce the 
the overall numerical factor, compared to PS theory, by a factor 
$1-\tilde{\kappa}$.  Note  that in the ST mass function 
the numerical value of the overall constant is fixed by hand, by
imposing the normalization condition that all the mass ends up in virialized objects. In our case, in contrast, the mass function comes out automatically with the correct normalization, as we already showed in Section~5.4 of  paper~I. The derivation given in eqs.~ (126)-(128) of paper~I goes through trivially when $\d_c\ra\d_c\sqrt{a}$, so the term proportional to the incomplete Gamma function  in \eq{ourf} ensures that the mass function is properly normalized, when the amplitude of the term proportional to 
$\exp\{-a\d_c^2/(2\s^2)\}$ is reduced by a factor $1-\tilde{\kappa}$.

\section{Conclusions}\label{sect:Concl}

In this paper we have proposed a generalization of excursion set theory, based on the idea that the threshold for collapse should be treated as a stochastic variable, fluctuating around the ellipsoidal collapse barrier (or, in the large mass limit, around the spherical collapse barrier). We have seen that fluctuations in the threshold arise naturally from a number of physical effects. For instance, even within the highly simplified description in which a halo is modeled 
as a smooth and homogeneous ellipsoid, fluctuations in the collapse barrier arise from the fact that the eigenvalues of its deformation tensor are stochastic variables, governed by a distribution probability. Only when one averages over this probability distribution one recovers a barrier which is a fixed function of the variance $\s^2(R)$ of the smoothed density field. Otherwise, as already discussed in
\cite{Audit,LeeS,SMT} one has a "fuzzy barrier" which  fluctuates
around the ellipsoidal collapse value, with fluctuations that can  
occasionally bring  the critical value for  collapse even  below the
spherical collapse value $\d_c$, see e.g. Fig.~\ref{fig:diff_barrier_3Sigma}.
As we have discussed, many other effects, such as the details of the halo finder, the environment, or the presence of non-linear substructures, contribute to these fluctuations.

For mathematical simplicity in this paper we have restricted ourselves
to 
a barrier that performs a diffusive motion, with diffusion constant
$D_B$, 
around the constant value
given by spherical collapse. We expect this to be a good approximation
in the large mass limit.
For such a barrier we have found that the first-passage time problem can be elegantly solved, and leads to a very simple result. Namely, in the mass function one must replace $\d_c\ra \sqrt{a}\d_c$, where $a=1/(1+D_B)$. The replacement
$\d_c\ra \sqrt{a}\d_c$
 is just the modification that was postulated in \cite{SMT}, in order
 to fit the results of $N$-body simulations.  The diffusing barrier
 model therefore offers a physical understanding of this modification of the PS mass function.

We have then combined our diffusing barrier model with the non-markovian corrections due to a tophat filter in coordinate space computed in paper~I, and we have presented an analytic expression for the halo mass function, valid for large masses. This result can be compared with existing 
$N$-body simulations. We have inferred the value of $D_B$ from the results presented in \cite{RKTZ}, and given $D_B$ our model predicts the corresponding value of $a$ for the same simulation. Our mass function, with $a$ fixed in this way,  is then compared to  the corresponding $N$-body simulations in  
Figs.~\ref{fig:compare_Nbody}--\ref{fig:compare_Nbody_ratio}.
The agreement is better than 20\% for all  $\sigma^{-1}\geq 0.3$ (corresponding 
approximately to halo masses $M\gsim 10^{11}\msun/h$) and  better than 10\% for all $\sigma^{-1}$ in the interval $1\leq\sigma^{-1}\leq 3$ (corresponding approximately to halo masses from $M\sim 10^{14}\msun/h$ to $M\sim 10^{15}\msun/h$).

We   conclude with an assessment of what can  be obtained from
excursion set theory, when it is combined with a diffusing barrier
model, and with a discussion of possible improvements of the model. First of all one should stress that
this theoretical model, using relatively simple ingredients, investigates a very complex phenomenon such as halo formation. It is therefore encouraging that it nevertheless provides an analytic result 
for the halo mass function that agrees with the $N$-body data with a precision better than $20\%$ over four decades  of halo masses (and the precision becomes of order 5-10\% in the higher mass range). Considering that over this mass range  the halo mass function changes by more than three orders of magnitude,  
this is a  non-trivial result.

We also stress that, when comparing our result to the $N$-body data as in Fig.~\ref{fig:compare_Nbody}, we 
had no freedom of adjusting  free parameters. The functional form of the halo mass function was derived from our model, and in this sense it has a different meaning compared to  many fitting  functions that have been proposed in the literature with the only aim of reproducing the $N$-body data.
We needed an input from $N$-body simulations, namely the 
scatter of the values of the threshold for collapse, from which we
deduced the
diffusion coefficient $D_B$ of the  barrier.
However, 
given this input our model makes a  prediction for the the numerical
value of the parameter $a$ that appears in our halo mass
function. This is  different from fitting $a$ directly to the $N$-body
simulations. The fact that the resulting halo mass function agrees
with $N$-body simulations much better than the original extended PS
theory lends support to the idea that 
the diffusing barrier model provides an effective way of including, within the excursion set theory framework, a number of physical effects that are lost when excursion set theory is combined with  the simpler spherical or ellipsoidal collapse models. 

For precision cosmology, especially in the future, an accuracy such as
the one that we have achieved is probably not yet sufficient. Of
course, generally speaking, analytic models are not meant to compete
with very time-consuming numerical simulations as far as accuracy is
concerned. Rather, their role is to provide some physical
understanding and some guidance.

However,  improvements of the model are certainly possible.
In particular,  rather than considering
a diffusive motion around the constant value $\d_c$, one should consider
the actual behavior of the threshold and of its variance with
$\s$. According to  \cite{RKTZ}, the average value of the barrier
basically follows the prediction (\ref{B(S)}) of the ellipsoidal
collapse, and the scatter around it has a
log-normal distribution. 
In general, fluctuations of the barrier below $\d_c$ are rare (the
vast majority of points in Fig.~3 of \cite{RKTZ} lie above $\d_c$,
since the average value follows the ellipsoidal collapse barrier,
which is a rising function of $\s$). However, as we discussed in
Section~\ref{sect:halo}, 
one should not forget that the fluctuations leading to massive clusters are rare events, which belong to the high-mass tail of the distribution 
function. Since 
the collapse  of a halo depends exponentially on the square of the
height of the barrier, even the rare occasional
fluctuations that bring the barrier below $\d_c$ can  end up enhancing significantly the formation probability of the rarest objects.
The first-passage time problem with such a
stochastic barrier might be hard to solve analytically, but one could
simply integrate numerically the corresponding Langevin equation, as
in \cite{Bond}. In the limit of small $\s$ one should recover the results
presented in this paper, but for intermediate values of $\s$ there
will be corrections.
We plan to investigate this issue in  future work.

Another possible future developement is the
 investigation of whether universality violations can be accounted for within the excursion set theory framework, supplemented by a stochastic barrier.
Recall that, within excursion set theory, the mass function can be written as in
\eq {dndMdef}, where the function $f$ depends only on the variance $\s$ of the smoothed density field. Thus, this function has a universal form in the sense that its
dependence on redshift and on cosmology enters only through the dependence of
the variance $\s(R,z)$.  In
$N$-body simulations  there are indications of violations of
universality, at approximately 10\%
level~\citep{Reed:2006rw,Tinker:2008ff}.\footnote{Observe however that
these violations of universality   depends  on the halo finder algorithm, and at least with some halo finders they can be accounted for  by systematic corrections due to the finite simulation volume \citep {Lukic:2007fc}.} The evolution with redshift of the exponential cutoff is minimal 
\citep{Tinker:2008ff} while a redshift dependence
shows up in the coefficients $a_T$, $b_T$ and $A_T$ of \eq{fitfs}.
Since $c_T$ and therefore our parameter $a$, do not show appreciable dependence with $z$, within our model these violations of universality cannot be ascribed  to a redshift dependence of
the diffusion coefficient $D_B$. However, $D_B$ only reflects the scatter of the barrier near $\s=0$.  In excursion set theory, the $\s$-dependent prefactors in front of the exponential, in the halo mass function, rather originate from  the shape of the ellipsoidal collapse barrier away from $\s=0$ \citep{SMT}. 
It would be interesting to study, with $N$-body simulations, the shape and the scatter of the collapse barrier as a function of the  redshift at which the halos eventually collapsed and virialized, i.e. to repeat for different $z$ the analysis performed in 
\cite{RKTZ} at $z=0$. One could then generate  an ensemble of
trajectories by integrating numerically
the corresponding Langevin equation, and study when these
trajectories  first  pierce such a stochastic
barrier. In this way one could investigate 
whether excursion set theory supplemented by  a
stochastic barrier can account for the observed deviations from
universality. The interest of such a procedure is that, if this were
indeed the case, one would have obtained 
some insight into the physical mechanisms responsible for the violations of universality. If, in contrast, this procedure should not reproduce the observed universality deviation, one would have to conclude that this is an intrinsic limitation of excursion set theory. In any case, one would get a better 
understanding of what can, and what cannot, be explained within the framework of this theoretical model.

Finally, another interesting test  of our model that could be performed with $N$-body simulations is the computation of the barrier scatter, and hence of $D_B$, with different halo finders (i.e. with different values of $\D$ in the spherical overdensity algorithm, and with different link-length in the FOF algorithm). Changing the halo finder changes  the exponential factor in the mass function, i.e. the constant $a$, and the diffusing barrier model predicts that the diffusion coefficient $D_B$ should change accordingly, in such a way that the relation $a=1/(1+D_B)$ is preserved.

\vspace{5mm}

\noindent We thank Sabino Matarrese, Ravi Sheth, Cristiano Porciani and Sidney Redner 
for useful discussions. The work
of MM is supported by the Fond National Suisse. 
The work of AR is supported by 
the European Community's Research Training Networks 
under contract MRTN-CT-2006-035505.

\appendix

\section{A. Inclusion of non-markovian corrections}

In this appendix we compute the effect of the non-markovian corrections due to a tophat filter in coordinate space.
As it was shown in paper I, when we use a tophat filter in coordinate
space the two-point correlation function can be written as
\be\label{rever}
\langle\delta(S_1)\delta(S_2)\rangle = 
{\rm min}(S_1,S_2) + \D(S_1,S_2)\, ,
\ee
where 
\be\label{Deltakappa}
\D(S_1,S_2)\simeq 
\kappa\frac{S_1(S_2-S_1)}{S_2}\, ,
\ee
and $\kappa\simeq 0.45$. The first term in the right-hand side of 
eq.~(\ref{rever}) is responsible for the
markovian contribution to the dynamics, and it
originates from a Dirac-delta  gaussian noise; 
the second term provides the non-markovian contribution. 
The reader is referred to paper I for more details. 

The fact that the barrier diffuses with a diffusion coefficient $D_B$
means that
\be\label{BB}
\langle B(S_1)B(S_2)\rangle = 
D_B\, {\rm min}(S_1,S_2) \, .
\ee
More generally, even the motion of the barrier can be subject to
non-markovian effects, so \eq{BB} should be generalized to
\be\label{BB2}
\langle B(S_1)B(S_2)\rangle = 
D_B{\rm min}(S_1,S_2)+\D_B(S_1,S_2) \, .
\ee
Making the rather natural assumption that
$\langle \d(S_1)B(S_2)\rangle=0$
and introducing the variable $X(S)=\d(S)-B(S)$, we see that
\be
\langle X(S_1)X(S_2)\rangle = 
(1+D_B) {\rm min}(S_1,S_2)+\D(S_1,S_2)+\D_B(S_1,S_2) \label{DSSDBSS}\, .
\ee
Thus, our problem becomes formally identical
to a problem for a single degrees of freedom $X(S)$, with
an absorbing boundary condition at $X=0$, with diffusion coefficient
$(1+D_B)$, and non-markovianities described by
$\D(S_1,S_2)+\D_B(S_1,S_2)$.

We now make the assumption that $\D_B(S_1,S_2)$ is small with respect
to $\D(S_1,S_2)$.
This assumption
could  be tested by extracting the correlator
$\langle B(S_1)B(S_2)\rangle$ from the $N$-body
simulations, similarly to how the variance
$\langle B^2(S)\rangle$ has been computed in \cite{RKTZ}. The
effect of a non-vanishing  $\D_B$ can  be
included perturbatively using the technique that we developed in
paper~I, just as we did for $\D(S_1,S_2)$. (Actually, we expect
that the two-point function
of the critical value for collapse $B(S)$  receives non-markovian 
corrections  due to the same smoothing procedure that gives the
non-markovian corrections to $\d(S)$ so, if this is the dominant
effect, a
plausible expectation is that
$\langle B(S_1)B(S_2)\rangle =
D_B [{\rm min}(S_1,S_2)+\D(S_1,S_2)]$, i.e. the barrier has the
same two-point function as $\d(S)$, apart from the overall diffusion
coefficient, so $\D_{B}(S_1,S_2)=D_B\D(S_1,S_2)$.
If this is the case, $\tilde{\kappa}$ in \eq{tildekappa} 
below is
replaced by $\kappa$. This would entail a ${\cal O}(25)\%$ modification of the
non-markovian correction computed below.)

When $\D_B$ can be neglected, the computation of the halo mass
function to first order in the non-markovian corrections can be
performed introducing
a rescaled ``time'' variable
$\tilde{S}=(1+D_B)S$. Then, using the explicit expression 
(\ref{Deltakappa}), we get
\be
\langle X(\tilde{S}_1)X(\tilde{S}_2)\rangle =
{\rm min}(\tilde{S}_1,\tilde{S}_2) +\tilde{\kappa}
\frac{\tilde{S}_1(\tilde{S}_2-\tilde{S}_1)}{\tilde{S}_2}
\ee
where
$\tilde{\kappa}=\frac{\kappa}{1+D_B}$
This is the same problem  that we have already solved in paper~I,
with $\kappa$ replaced by $\tilde{\kappa}$ and 
$S$ replaced by $\tilde{S}$, so the
solution can  be written immediately, and is given by \eqs{ourf}{a79}.


\begin{thebibliography}{99}
\expandafter\ifx\csname natexlab\endcsname\relax\def\natexlab#1{#1}\fi

\bibitem[Audit {et.~al.}(1997)]{Audit}
Audit, E., Teyssier, R. and Alimi, J.-M., 1997,
Astron. Astrophys. 325,439.

\bibitem[Bardeen  {et~al.}(1986)]{BBKS}
  Bardeen, J.~M., Bond, J.~R.,Kaiser N. and Szalay, A.~S., 1986,
  ApJ  {304}, 15.

\bibitem[Bond {et~al.}(1991)]{Bond}
  Bond, J.~R., Cole, S., Efstathiou, G. \& Kaiser, N. 1991,
  ApJ.  { 379},  440.

\bibitem[Bond \& Myers(1996)]{BondMyers} 
Bond, J.~R. and Myers, S. 1996, ApJS, 103, 1.

\bibitem[Dalal et al.(2008)]{Dalal:2008zd}
  Dalal, N., White, M., Bond, J.~R. \& Shirokov, A.
  arXiv:0803.3453 [astro-ph].
 
 \bibitem[Desjacques(2008)]{Desjacques}
  Desjacques, V.~,
  MNRAS 388,  638.

 \bibitem[Doroshkevich(1970)]{Dorosk} Doroshkevich, A.~G. 1970,
Astrofizika, 3,175.


\bibitem[\protect\citeauthoryear{Grinstein \& Wise}{1986}]{GW} Grinstein, B.,
  \& Wise, M.~B.\ 1986, ApJ, 310, 19. 

\bibitem[Grossi et al.(2009)]{grossi2009} 
Grossi, M. et al., arXiv:0902.2013.


\bibitem[H\"{a}nggi et al.(1981)]{hanggi1981}
H\"{a}nggi, P. 1981, Z. Phys. {B45}, 79.

\bibitem[Katz et al.(1993)]{KQG} Katz, N., Quinn, T. \& Gelb,~J.~M. 1993,
\mnras, 265, 689.

\bibitem[Koyama et al.(1999)]{KOYAMA} Koyama, K., Soda,
J., \& Taruya, A.\ 1999, MNRAS, 310, 1111. 

\bibitem[Knessl(1986)]{knessl} 
Knessl, C. {\em et al.} 1986, J. Stat. Phys. { 42},  169.

\bibitem[Jeltema et al.(2005)]{Jeltema}
Jeltema, ~T. ~E., Canizares,~C.~R., Bautz,~M.~W., Buote,~D.~A,
ApJ 624, 606

\bibitem[Jenkins et al.(2001)]{jenkins} Jenkins, A.  
{\em et al.} 2001, MNRAS {321}, 372.

\bibitem[Lam et al.(2009)]{Lam:2009nd}
  Lam, T.~Y., Sheth, R.~K. \& Desjacques, V.,
  arXiv:0905.1706 [astro-ph.CO].

\bibitem[Lee \& Shandarin(1998)]{LeeS} 
Lee, J. \& Shandarin,~S.~F. 1998, ApJ, 500, 14.


\bibitem[\protect\citeauthoryear{Lucchin et al.}{1988}]{LMV} Lucchin, F.,
Matarrese, 
S., \& Vittorio, N.\ 1988, ApJl, 330, L21. 

\bibitem [Lukic et al(2007)]{Lukic:2007fc}
  Lukic,  Z.~, Heitmann, K.~, Habib, S.~, Bashinsky, S.~ and Ricker, P.~M. 2007,
  ApJ 671, 1160.

\bibitem[Lukic et al.(2009)]{Lukic:2008ds}
  Lukic, Z.~, Reed, D.~, Habib, S.~ \& Heitmann, K. 2009,
  ApJ, 692, 217.


\bibitem[Maggiore \& Riotto(2009a)]{MR1} Maggiore,  M. and Riotto, A., 
	arXiv:0903.1249 [astro-ph.CO],  (paper I).

\bibitem[Maggiore \& Riotto(2009c)]{MR3} Maggiore,  M. and Riotto, A., 
	arXiv:0903.1251 [astro-ph.CO], (paper III).

\bibitem[Matarrese et al.(1986)]{MLB} Matarrese, S., 
Lucchin, F., \& Bonometto, S.~A.\ 1986, ApJ., 310, L21.

\bibitem[{Matarrese} {et~al.}(2000)]{MVJ}
  Matarrese, S., Verde, L.  \& Jimenez,~R. 2000,
  ApJ { 541}, 10.

\bibitem[{Moscardini et al.}(1991)]{MMLM} Moscardini,
L.,  Matarrese, S., Lucchin, F., \& Messina, A.\ 1991, MNRAS, 248, 424. 


\bibitem[Pillepich {et~al.}(2008)]{PPH}
Pillepich, A. Porciani, C.  \& Hahn, O.  2008,
 arXiv:0811.4176 [astro-ph].

\bibitem[Press \& Schechter(1974)]{PS} 
Press,  W. H. \& Schechter,~P. 1974,
ApJ { 187},  425.

\bibitem[Redner(2001)]{redner2001}
Redner, S. 2001,
``A guide to first-passage processes" (Cambridge University
press).

\bibitem[Reed et al.(2006)]{Reed:2006rw}
  Reed, D.~, Bower, R.~, Frenk,  C.~,  Jenkins, A.~  \& Theuns, T. 2007,
  MNRAS 374,  2.


\bibitem[Robertson et al.(2008)]{RKTZ}
  Robertson, B. et al. 2008,
  arXiv:0812.3148 [astro-ph].
 
\bibitem[Robinson \& Baker(2000)]{RB} Robinson, J., \&
Baker, J.~E.\ 2000, MNRAS, 311, 781. 

\bibitem[Robinson et al.(2000)]{RGS} Robinson, J.,
Gawiser, E., \& Silk, J.\ 2000, ApJ, 532, 1. 


\bibitem[Sandvik et al.(2006)]{Sandvik}
  Sandvik, H.~B., Moeller, O.~, Lee, J.~ and White, S.~D.~M. 2006,
  MNRAS, 377, 234.

\bibitem[Sheth et al.(2001)]{SMT} 
Sheth, R.~K., Mo, H.~J., \& Tormen, G.\ 2001, \mnras, 323, 1.

\bibitem[Sheth \& Tormen(1999)]{ST} 
Sheth, R.~K., \& Tormen, G.\ 1999, \mnras, 308, 119. 

\bibitem[Springel et al.(2005)]{Springel:2005nw}
  Springel, V. {\it et al.} 2005,
  Nature {435}, 629
 
\bibitem[Sugiyama(1995)]{Sugiyama} Sugiyama, N. 1995, ApJS 100, 281.

\bibitem[Tinker {et~al.}(2008)]{Tinker:2008ff}
  Tinker J.~L.~{\it et al.} 2008,
  ApJ, 688, 709.


\bibitem[van Kampen \& Oppenheim(1972)]{vKampen} 
van Kampen N. G. \&  Oppenheim, I. 1972,
J. Math. Phys. { 13}  842.

\bibitem[van Kampen(1998)]{vkampen1998}
van Kampen N. G., 1998, Braz. Journ. of Phys. {28} 90.


\bibitem[Warren {et~al.}(2006)]{Warren:2005ey}
  Warren, M.~S., Abazajian, K.,  Holz,D.~E. \& Teodoro, L. 2006,
  ApJ{ 646}  881.

\bibitem[Weiss  {et~al.}(1983)]{weiss1983}
Weiss G. H.  et al. 1983, 
Physica {119A} 569.

\bibitem[White(2001)]{White}
White, M. 2001,
Astron. Astrophys. 367, 27.

\bibitem[Zentner(2007)]{Zentner}
  Zentner, A.~R. 2007,
  Int.\ J.\ Mod.\ Phys.\  D { 16}  763.


\end{thebibliography}
\end{document}